\documentclass[conference,comsoc,twocolumn,twoside,10pt]{IEEEtran}
\usepackage{amsmath,graphicx,subfig}
\usepackage{caption}
\usepackage{enumitem}
\usepackage{amsmath}
\usepackage{mathtools,amssymb}
\usepackage{romannum}
\usepackage{algpseudocode}
\usepackage{algorithm}

\begin{document}
\title{Study of SIC and RLS Channel Estimation for Large-Scale Antenna Systems with 1-Bit ADCs}

\author{Zhichao Shao, Lukas~T.~N.~Landau and Rodrigo C. de Lamare \\
    Center for Telecommunications Studies (CETUC) \\ Pontifical
    Catholic University of Rio de Janeiro, RJ, Brazil.\\
    Emails: \{zhichao.shao;lukas.landau;delamare\}@cetuc.puc-rio.br
    }

\maketitle

\begin{abstract}

We propose a novel low-resolution-aware recursive least squares
channel estimation algorithm for uplink multi-user multiple-input
multiple-output systems. In order to reduce the energy consumption,
1-bit ADCs are used on each receive antenna. The loss of performance
can be recovered by the large-scale antenna arrays at the receiver.
The proposed adaptive channel estimator can mitigate the distortions
due to the coarse quantization. Moreover, we propose a
low-resolution-aware minimum mean square error based successive
interference canceler to successively mitigate the multiuser
interference. Simulation results show good performance of the system
in terms of mean square error and bit error rate.
\end{abstract}

\begin{IEEEkeywords}
Large-scale multiple-antenna systems, 1-bit quantization, recursive least squares, minimum mean square error, successive interference cancellation
\end{IEEEkeywords}

\section{Introduction}

In the next generation communication systems, large-scale antenna
systems at the base station (BS) have been identified as a key
technique due to its high spectral and energy efficiency
\cite{Larsson}. However, with a large number of antennas at BS the
energy consumption will increase drastically. With an abundance of
spatial degrees of freedoms (DoFs), the receiver can be designed
with many inexpensive low-resolution components. In fact, 1-bit
analog-to-digital converters (ADCs) at the front-end can
dramatically decrease the receiver energy consumption. However, due
to the large distortions of the coarse quantization the standard
estimation and detection methods often yield poor performances.
Recently, there have been some works studying channel estimation and
detection algorithms in 1-bit ADCs systems \cite{risi,Li,6196608}.
The authors in \cite{risi} have made an investigation about massive
multiple-input multiple-output (MIMO) systems with 1-bit ADCs, where
they have used least squares (LS) channel estimation, maximal ratio
combining (MRC) and zero-forcing (ZF) detection. In \cite{Li} the
authors have proposed another channel estimator, named Bussgang
linear minimum mean squared error (BLMMSE), which can achieve better
mean square error (MSE) performance but with high computational
cost.

In the last decade, low-density parity-check (LDPC) codes
\cite{7489038} have been widely used in many industry standards
including WiMAX, WiFi and DVB-S2. More recently, they have also been
adopted for the next generation communication systems, since they
can approach the Shannon capacity with low complexity. To the best
of our knowledge, most of the previous works on 1-bit massive MIMO
mainly focus on the uncoded systems. It is also interesting to study
the system with channel coding. For this reason, we investigate the
1-bit large-scale antenna systems using LDPC codes.

In this paper, we devise an adaptive pilot-aided channel estimator,
named low-resolution-aware recursive least squares (LRA-RLS), that
operates on systems with  1-bit ADCs. Compared to previous
estimators, the proposed LRA-RLS algorithm can achieve high accuracy
but with low computational cost. For further performance
improvement, we develop a low-resolution-aware (LRA) minimum mean
square error (MMSE) based successive interference cancellation (SIC)
technique. It decodes different users sequentially, i.e., the
interference due to the decoded users is subtracted before decoding
other users.

The rest of this paper is organized as follows: Section \Romannum{2}
shows the system model and presents some statistical properties
about 1-bit quantization. Section \Romannum{3} illustrates the
derivation of the proposed LRA-RLS channel estimation. Section
\Romannum{4} describes the proposed LRA-MMSE-SIC technique. Section
\Romannum{5} gives the simulation results and Section \Romannum{6}
gives a summary of the work.

Notation: Matrices are represented by bold capital letters while
vectors are in bold lowercase. $\mathbf{I}_n$ denotes an $n\times n$
identity matrix. $\mathbf{0}_n$ is a $n\times 1$ all zeros column
vector. Additionally, diag$(\mathbf{A})$ is a diagonal matrix only
containing the diagonal elements of $\mathbf{A}$. The complex
conjugate is represented by $(.)^*$.

\section{System Model and Statistical Properties of 1-bit Quantization}

In this paper a single-cell multi-user large-scale multiple-antenna
scenario is considered, which is shown in Fig.
\ref{fig:transmitter}. $K$ single-antenna terminals are
simultaneously transmitting signals to the BS equipped with $M$
receive antennas, where $M\gg K$. The information bits $b_k$ are
encoded by the encoder and modulated to $x_k$ according to a
modulation scheme. The received unquantized signal $\mathbf{y}$ is
given by
\begin{equation}
\mathbf{y}=\mathbf{H}\mathbf{x}+\mathbf{n}=\sum_{k=1}^{K}\mathbf{h}_kx_k+\mathbf{n},
\end{equation}
where $\mathbf{H}\in \mathbb{C}^{M\times K}$ is the channel matrix
and $x_k$ is the transmit symbol with energy $\sigma_x^2$. The noise
vector $\mathbf{n}\sim
\mathcal{CN}(\mathbf{0}_M,\sigma^2_n\mathbf{I}_M)$ contains complex
Gaussian noise samples with zero mean and variance $\sigma^2_n$.

Let $\mathcal{Q}(.)$ represents the 1-bit quantization at the
receiver, the resulting quantized signal $\mathbf{y}_\mathcal{Q}$ is
\begin{equation}
\mathbf{y}_\mathcal{Q}
=\mathcal{Q}(\mathbf{y})=\mathcal{Q}\left(\mathfrak{R}\{\mathbf{y}\}\right)
+ j\mathcal{Q}\left(\mathfrak{I}\{\mathbf{y}\}\right),
\label{equ_system_model_quantize}
\end{equation}
where $\mathfrak{R}\{\cdot\}$ and $\mathfrak{I}\{\cdot\}$ get the
real and imaginary part, respectively. They are element-wisely
quantized to $\{\pm\frac{1}{\sqrt{2}}\}$ based on a threshold.

The cross-correlation between the unquantized complex Gaussian
distributed signal $\mathbf{s}$ with covariance matrix
$\mathbf{C}_\mathbf{s}$ and its 1-bit quantized signal
$\mathbf{s}_\mathcal{Q}$ is described by \cite{Bussgang}
\begin{equation}
\mathbf{C}_{\mathbf{s}\mathbf{s}_\mathcal{Q}}=\sqrt{\frac{2}{\pi}}\mathbf{K}\mathbf{C}_{\mathbf{s}},\mbox{where } \mathbf{K}=\text{diag}\left(\mathbf{C}_{\mathbf{s}}\right)^{-\frac{1}{2}}.
\label{equ_bussgang}
\end{equation}
Furthermore, the covariance matrix of the 1-bit quantized signal
$\mathbf{s}_\mathcal{Q}$ is given by \cite{Jacovitti}
\begin{equation}
\mathbf{C}_{\mathbf{s}_\mathcal{Q}}=\frac{2}{\pi}\left(\text{sin}^{-1}\left(\mathbf{K}\mathfrak{R}\{\mathbf{C}_{\mathbf{s}}\}\mathbf{K}\right)+j\text{sin}^{-1}\left(\mathbf{K}\mathfrak{I}\{\mathbf{C}_{\mathbf{s}}\}\mathbf{K}\right)\right).
\label{equ_Jacovitti}
\end{equation}
Based on (\ref{equ_bussgang}) and (\ref{equ_Jacovitti}), we can
reformulate (\ref{equ_system_model_quantize}) as a statistically
equivalent linear system, which is given by
\begin{equation}
\mathbf{y}_\mathcal{Q}=\mathbf{Ay}+\mathbf{n}_q,
\label{fig:system_model}
\end{equation}
where $\mathbf{A}$ is the linear operator and chosen independently
from $\mathbf{x}$ and $\mathbf{n}$. The vector $\mathbf{n}_q$
contains the statistically equivalent quantizer noise with
covariance matrix
$\mathbf{C}_{\mathbf{n}_q}=\mathbf{C}_{\mathbf{y}_\mathcal{Q}}-\mathbf{A}\mathbf{C}_{\mathbf{y}}\mathbf{A}^H$.
The matrix $\mathbf{A}$ is calculated as
\begin{equation}
\mathbf{A}=\mathbf{C}_{\mathbf{yy}_\mathcal{Q}}^H\mathbf{C}_{\mathbf{y}}^{-1}=\sqrt{\frac{2}{\pi}}\text{diag}\left(\mathbf{C}_{\mathbf{y}}\right)^{-\frac{1}{2}},
\end{equation}
where $\mathbf{C}_{\mathbf{y}}$ denotes the auto-correlation matrix
of $\mathbf{y}$.

\section{Proposed LRA-RLS Channel Estimation}

Compared to LMS-based algorithms, the RLS algorithms can achieve
fast convergence and excellent performance when working in
time-varying environments for multiple-antenna systems
\cite{6612905,jidf}. In the uplink, each transmission packet
contains pilots and data symbols. During the training phase, all $K$
users simultaneously transmit $\tau$ pilots to the receiver, which
yields
\begin{equation}
\mathbf{Y}_{\mathcal{Q}_p}=\mathcal{Q}\left(\mathbf{Y}_p\right)=\mathcal{Q}\left(\mathbf{H}\mathbf{X}_p+\mathbf{N}_p\right),
\end{equation}
where $\mathbf{X}_p\in \mathbb{C}^{K\times \tau}$ is the pilot matrix, in which the $k$th row represents the transmitted pilots at the $k$th terminal. The received signal at the $m$th receive antenna is given by
\begin{equation}
\mathbf{y}^{m}_{\mathcal{Q}_p} = \mathcal{Q}\left(\mathbf{X}^H_p\mathbf{h}^{m}+\mathbf{n}_p^{m}\right),
\end{equation}
with $\mathbf{y}^m_{\mathcal{Q}_p}=\left[y^m_{\mathcal{Q}_p}(1),y^m_{\mathcal{Q}_p}(2),...,y^m_{\mathcal{Q}_p}(\tau)\right]^H$ and $\mathbf{n}^m_{p}\in \mathbb{C}^{\tau\times 1}$. The proposed LRA-RLS algorithm can be derived by solving the following least-squares optimization problem:
\begin{equation}
\begin{aligned}
\hat{\mathbf{h}}^m(n)=\text{arg} \min_{\mathbf{h}^m(n)}&\sum_{n=1}^{\tau}\lambda^{\tau-n}\left|{y}^{m}_{\mathcal{Q}_p}(n)-A_p(n)\mathbf{h}^m(n)^H\mathbf{x}_p(n)\right|^2\\&\hspace{5mm}+\delta\lambda^{\tau}\left|\left|\mathbf{h}^m(n)\right|\right|^2_2,
\end{aligned}
\end{equation}
where $\lambda$ is the forgetting factor and $\delta$ is the regularization
factor. $\mathbf{x}_p(n)=\left[x_{1_p}(n),x_{2_p}(n),...,x_{k_p}(n)\right]^T$. From (6), the linear operator is chosen as
\begin{equation}
A_p(n)=\mathbf{C}^H_{\mathbf{y}^m_p(n)\mathbf{y}^m_{\mathcal{Q}_p}(n)}\mathbf{C}^{-1}_{\mathbf{y}^m_p(n)}= \sqrt{\frac{2}{\pi}}\left(\mathbf{x}_p(n)^H\mathbf{x}_p(n)+\sigma_n^2\right)^{-\frac{1}{2}}.
\end{equation}
The proposed channel estimator is described in Algorithm \ref{Proposed_RLS}.

\section{Proposed Low-resolution-aware Detection Schemes}

In this section, we describe a linear LRA-MMSE detection scheme
\cite{8240730} and propose a LRA-MMSE-SIC detector, which employs
the LRA-MMSE filter and performs the successive interference
cancellation technique at the receiver.

\subsection{Proposed Linear LRA-MMSE detector}

Recall the system model in (\ref{fig:system_model}), the linear
LRA-MMSE filter $\mathbf{W}$ is applied to
$\mathbf{y}_{\mathcal{Q}}$, to obtain
\begin{equation}
\hat{\mathbf{x}}=\mathbf{W}^H\mathbf{y}_{\mathcal{Q}},
\label{estimated_symbol}
\end{equation}
where $\mathbf{W}$ is chosen to minimize the MSE between the transmitted symbol $\mathbf{x}$ and the filter output, i.e.
\begin{equation}
\mathbf{W}=\arg\min_{\mathbf{W'}} E\left[\left\vert\left\vert \mathbf{x}-\mathbf{W'}^{H}\mathbf{y}_{\mathcal{Q}}\right\vert\right\vert^2_2\right].
\end{equation}
The solution is given by
\begin{equation}
\mathbf{W}=\mathbf{C}_{\mathbf{y}_{\mathcal{Q}}}^{-1}\mathbf{C}_{\mathbf{y}_{\mathcal{Q}}\mathbf{x}},
\end{equation}
where the covariance matrix is
\begin{equation}
\mathbf{C}_{\mathbf{y}_\mathcal{Q}}=\frac{2}{\pi}\left(\text{sin}^{-1}\left(\mathbf{K}\mathfrak{R}\{\mathbf{C}_{\mathbf{y}}\}\mathbf{K}\right)+j\text{sin}^{-1}\left(\mathbf{K}\mathfrak{I}\{\mathbf{C}_{\mathbf{y}}\}\mathbf{K}\right)\right)
\end{equation}
and the cross-correlation vector is
\begin{equation}
\mathbf{C}_{\mathbf{y}_{\mathcal{Q}}\mathbf{x}}=\sigma_x^2\mathbf{A}\mathbf{H}.
\end{equation}
Note that $\mathbf{C}_{\mathbf{y}}$ is the covariance matrix of the unquantized data vector $\mathbf{y}$, which leads to the following result:
\begin{equation}
\mathbf{C}_{\mathbf{y}}=E\left[\left(\mathbf{H}\mathbf{x}+\mathbf{n}\right)\left(\mathbf{H}\mathbf{x}+\mathbf{n}\right)^H\right]=\sigma_x^2\mathbf{HH}^H+\sigma_n^2\mathbf{I}_M.
\end{equation}

\subsection{Proposed LRA-MMSE-SIC hard detector}

The idea of SIC for multiple-antenna systems has been first
introduced in \cite{Wolniansky} and can be considered one of the
most efficient IC-based techniques \cite{spa,PLi}. It removes the
interference in a recursive way: the interference imposed by
previous particular symbol on the current to be detected symbol is
subtracted after recreating the interference. The system model here
is slightly different to Fig. \ref{fig:transmitter}. The output of
the detector can be directly sent to the demodulator without the
blocks of decoder and hard decision. The proposed algorithm
including the ordering is illustrated in Algorithm \ref{alg:sic}.

In (\Romannum{2}.1) $(\mathbf{G}_i)_{k_i}$ represents the ${k_i}$th column of $\mathbf{G}_i$. The operation \textbf{Mod}(.) in (\Romannum{2}.3) denotes the slicing
operation appropriate to one of the complex constellation set with $2^{M_c}$ possible points. Thus, (\Romannum{2}.4) performs interference cancellation of the detected symbol from the received vector; (\Romannum{2}.5) assigns an all-zeros column vector to the $k_i$th column of channel matrix $\mathbf{H}$; (\Romannum{2}.6)-(\Romannum{2}.9) compute the receive filter for the next iteration; (\Romannum{2}.10) determines the next optimal ordering. Note that the columns $k_1,k_2,\dots,k_i$ in $\mathbf{G}_{i+1}$ have been zeroed, because the corresponding interferences have been canceled.

\subsection{Proposed LRA-MMSE-SIC soft detector}
Based on the previous proposed hard detector and existing iterative detection and decoding schemes \cite{8240730,7105928}, we derive here a soft detector, which produces soft information to the channel decoder. In Algorithm \ref{alg:sic}, (\Romannum{2}.3) can be replaced with the following steps.

In (\Romannum{2}.2) instead of a recursive way, $\mathbf{y}_{\mathcal{Q}_{k_{i}}}$ can also be calculated as
\begin{equation}
\begin{aligned}
\mathbf{y}_{\mathcal{Q}_{k_{i}}}&=\mathbf{y}_{\mathcal{Q}}-\mathbf{A}\left(\sum_{j=1}^{i-1}(\mathbf{H})_{k_{j}}\tilde{x}_{k_{j}}\right)\\&=\mathbf{A}\left(\mathbf{Hx}-\sum_{j=1}^{i-1}(\mathbf{H})_{k_{j}}\tilde{x}_{k_{j}}\right)+\mathbf{An}+\mathbf{n}_q
\end{aligned}
\end{equation}


In order to calculate $P(\tilde{x}_{k_i}|x_{k_i})$, we use the Cramer's central limit theorem \cite{cramer}: the LRA-MMSE filter output can be approximated by a complex Gaussian distribution due to the large number of independent variables with $\tilde{x}_{k_i}=\mu_{k_i}x_{k_i}+z_{k_i}$, where
\begin{equation}
\mu_{k_i}=\mathbf{w}_{k_i}^H\mathbf{A}(\mathbf{H})_{k_{i}}
\end{equation}
and $z_{k_i}$ is a zero-mean complex Gaussian variable with variance $\eta_{k_i}^2$ given by
\begin{equation}
\eta_{k_i}^2 =\sigma_x^2(\mu_{k_i} - \mu_{k_i}^2).
\end{equation}
Therefore, the likelihood function can be approximated by
\begin{equation}
P(\tilde{x}_{k_i}|x_{k_i})\simeq\frac{1}{\pi\eta_{k_i}^2}\text{exp}\left(-\frac{1}{\eta_{k_i}^2}\left\vert\tilde{x}_{k_i}-\mu_{k_i}x_{k_i}\right\vert^2\right).
\end{equation}
Then the log likelihood ratio (LLR) computed by the LRA-MMSE filter for the $l$th bit ($l\in \{1,...,M_c\}$) of the symbol $\tilde{x}_{k_i}$ is given by
\begin{equation}
\begin{aligned}
\log\frac{\text{Pr}\left(\tilde{x}_{k_i}\vert b^{l}_{k_i}=+1\right)}{\text{Pr}\left(\tilde{x}_{k_i}\vert b^{l}_{k_i}=-1\right)}&=\log\frac{\text{Pr}\left(b^{l}_{k_i}=+1\vert\tilde{x}_{k_i}\right)}{\text{Pr}\left(b^{l}_{k_i}=-1\vert\tilde{x}_{k_i}\right)}-\log\frac{\text{Pr}\left(b^{l}_{k_i}=+1\right)}{\text{Pr}\left(b^{l}_{k_i}=-1\right)}\\&=\log\frac{\sum_{x_{k_i}\in \mathcal{A}^{+1}_l} P\left(\tilde{x}_{k_i}|x_{k_i}\right)\text{Pr}\left(x_{k_i}\right)}{\sum_{x_{k_i}\in \mathcal{A}^{-1}_l} P\left(\tilde{x}_{k_i}|x_{k_i}\right)\text{Pr}\left(x_{k_i}\right)},
\end{aligned}
\end{equation}
where $\mathcal{A}^{+1}_l$ is the set of hypotheses $x_{k_i}$ for which the $l$th bit is +1 and $\mathcal{A}^{-1}_l$ is similarly defined. The a priori LLR is assumed to be zero with the assumption $\text{Pr}(b^{l}_{k_i}=+1)\simeq\text{Pr}(b^{l}_{k_i}=-1)$ and the a priori probabilities $\text{Pr}(x_{k_i})$ are supposed to be the same without a priori information.

\section{Numerical Results}

In this section, we evaluate our proposed LRA-RLS channel estimator
and the LRA-MMSE-SIC detector in terms of the normalized MSE and bit
error rate (BER) and compare them with current existing estimators
and detectors. The modulation scheme is QPSK, where $\sigma_x^2 =
1$. The channel is assumed to experience block fading. In the
LRA-RLS channel estimation phase $\lambda$ is chosen as 0.94. The
value of $\delta$ is not fixed and it increases from $10^{-11}$ to
$3\times10^{-1}$ while the SNR grows. For the coded system, we
consider a short length regular LDPC code with block length 512 and
rate 1/2. The decoding method used in channel decoder is the sum
product algorithm (SPA).

The normalized MSE performance of the proposed channel estimator and
other estimators is depicted in Fig. 2, which shows the proposed
LRA-RLS channel estimator ($\mathcal{O}(K^2(M\tau))$) achieves a
close performance to that of the BLMMSE algorithm
($\mathcal{O}((M\tau)^3)$) but with lower complexity. The uncoded
BER performance illustrated in Fig. 3 indicates that the proposed
LRA-MMSE-SIC hard detector outperforms the other existing detectors.
In Fig. \ref{fig:coded_BER} the LDPC coded BER is simulated as
function of the $E_b/N_0$ under the perfect channel state
information (CSI) and the BLMMSE estimated CSI. It can be seen from
the results that the LRA-MMSE-SIC soft detector leads to the best
BER performance. Moreover, for considering the dynamic range of the
receiver in terms of handling terminals with different receive power
levels at the same time, we consider the scenario where one terminal
has a higher transmit energy than the others. The BER performance of
all terminals is shown in Fig. \ref{fig:sensitivity}, where $n$dB
means the transmit energy of the terminal with the highest transmit
energy is $n$dB higher than the transmit energy of the others.

%
%
%

\section{Conclusion}

This work has proposed an adaptive RLS channel estimator for 1-bit
large-scale antenna systems. The simulation results have shown good
MSE performance with low computational cost. Furthermore, LRA-MMSE
based hard and soft SIC detector have been developed for the 1-bit
systems. They provide the best BER performances among the detectors
in \cite{risi}.

\bibliographystyle{IEEEtran}
\bibliography{ref}

\end{document}